\documentclass{aastex}
\usepackage{emulateapj5}
\usepackage{graphicx}
\usepackage{times,psfig}

 \usepackage{url}
\def\spose#1{\hbox to 0pt{#1\hss}}
\newcommand\lsim{\mathrel{\spose{\lower 3pt\hbox{$\mathchar"218$}}
     \raise 2.0pt\hbox{$\mathchar"13C$}}}
\newcommand\gsim{\mathrel{\spose{\lower 3pt\hbox{$\mathchar"218$}}
     \raise 2.0pt\hbox{$\mathchar"13E$}}}
\def\ltsima{$\; \buildrel < \over \sim \;$}
\def\lsim{\lower.5ex\hbox{\ltsima}}
\def\gtsima{$\; \buildrel > \over \sim \;$}
\def\gsim{\lower.5ex\hbox{\gtsima}}

\def\ergcms{{\rm\thinspace erg \thinspace cm^{-2} \thinspace s^{-1}}}

\shorttitle{High redshift blazars observed by {\it NuSTAR}}
\shortauthors{Tagliaferri G., et al.}

\begin{document}

\title{{\it NuSTAR} and multifrequency study of the two high--redshift blazars S5 0836+710 and PKS 2149--306}

\author{G. Tagliaferri\altaffilmark{1},  G. Ghisellini\altaffilmark{1}, M. Perri\altaffilmark{2,3}, M. Hayashida\altaffilmark{4}, M. Balokovi\'c\altaffilmark{5}, S. Covino\altaffilmark{1}, P. Giommi\altaffilmark{2},  G.M. Madejski\altaffilmark{6}, S. Puccetti\altaffilmark{2,3}, T. Sbarrato\altaffilmark{1,7}, S.\ E.\ Boggs\altaffilmark{8}, J. Chiang\altaffilmark{6}, F.\ E.\ Christensen\altaffilmark{9}, W.W. Craig\altaffilmark{9,10}, C.\ J.\ Hailey\altaffilmark{11}, F.\ A.\ Harrison\altaffilmark{5}, D.\ Stern\altaffilmark{12}, W.\ W.\ Zhang\altaffilmark{13}}

\altaffiltext{1}{INAF -- Osservatorio Astronomico di Brera, via E. Bianchi 46, I--23807 Merate, Italy; gianpiero.tagliaferri@brera.inaf.it}
\altaffiltext{2}{ASI -- Science Data Center, via del Politecnico, I-00133 Rome, Italy}
\altaffiltext{3}{INAF -- Osservatorio Astronomico di Roma, via Frascati 33, I--00040 Monteporzio Catone, Italy}
\altaffiltext{4}{Institute for Cosmic Ray Research, University of Tokyo, 5-1-5 Kashiwanoha, Kashiwa, Chiba, 277-8582, Japan}
\altaffiltext{5}{Cahill Center for Astrophysics, California Institute of Technology, 1200 East California Boulevard, Pasadena, CA 91125, USA}
\altaffiltext{6}{Kavli Institute for Particle Astrophysics and Cosmology, SLAC National Accelerator Laboratory, Menlo Park, CA 94025}
\altaffiltext{7}{Dipartimento di Fisica G.\ Occhialini, Universit\`a di Milano Bicocca, Piazza della Scienza 3, I--20126 Milano, Italy}
\altaffiltext{8}{Space Sciences Laboratory, University of California, Berkeley, CA 94720}
\altaffiltext{9}{DTU Space - National Space Institute, Technical University of Denmark, Elektrovej 327, 2800 Lyngby, Denmark}
\altaffiltext{10}{Lawrence Livermore National Laboratory, Livermore, CA 94550, USA}
\altaffiltext{11}{Columbia Astrophysics Laboratory, Columbia University, New York, NY 10027} 
\altaffiltext{12}{Jet Propulsion Laboratory, California Institute of Technology, Pasadena, CA 91109}
\altaffiltext{13}{NASA Goddard Space Flight Center, Greenbelt, MD 20771}

\received{~~} \accepted{~~}
\journalid{}{}
\articleid{}{}
\authoremail{gianpiero.tagliaferri@brera.inaf.it}

\begin{abstract}
The most powerful blazars are the flat spectrum radio quasars whose emission is dominated
by a Compton component peaking between a few hundred keV and a few hundred MeV.
We selected two bright blazars, PKS~2149--306 at redshift $z$=2.345 and S5~0836+710 at $z$=2.172,
in order to observe them in the hard X-ray band with the {\it NuSTAR} satellite. 
In this band the Compton component is rapidly rising almost up to the peak of the emission. 
Simultaneous soft-X-rays and UV-optical observations were performed with the {\it Swift} 
satellite, while near-infrared (NIR) data were obtained with the REM telescope.
To study their variability, we repeated these observations for both sources on a timescale of
a few months. 
While no fast variability was detected during a single observation, both sources
were found to be variable in the X-ray band, up to 50\%, between the two observations, 
with larger variability at higher energies. 
No variability was detected in the optical/NIR band. 
These data together with {\it Fermi}--LAT,
{\it WISE} and other literature data are then used to study the overall 
spectral energy distributions (SEDs) of these blazars.
Although the jet non-thermal emission dominates the SED, it leaves the UV band unhidden,
allowing us to detect the thermal emission of the disc and to estimate the mass of the black hole.
The non-thermal emission is well reproduced by a one-zone leptonic model.
The non-thermal radiative processes are synchrotron, self-Compton and 
external Compton using seed photons from both the broad-line region (BLR) and the torus. 
We find that our data are better reproduced if we assume
that the location of the dissipation region of the jet, $R_{\rm diss}$, is in-between
the torus, (at $R_{\rm torus}$), and the BLR ($R_{\rm torus}>R_{\rm diss}>R_{\rm BLR}$).
The observed variability is explained by changing a minimum number of model parameters by 
a very small amount.
\end{abstract}

\keywords{galaxies: active -- quasars: general -- X-rays: general --- 
		blazar: individual: PKS 2149--306, S5 0836+710}


\section{Introduction}
\label{sec-intro}

Blazars are a sub-class of AGN whose emission is dominated by a relativistic jet pointing toward us.
Their emission extends from the radio band to the $\gamma$-ray and TeV band,
and is dominated by non-thermal processes. 
Their spectral energy distribution (SED) is dominated by two humps. 
The first, spanning the infrared to the X-ray band, 
is usually attributed to synchrotron emission, while the second, located from the X-ray to 
the $\gamma$-ray band, is attributed to the inverse Compton scattering process.
This second component can be produced by energetic electrons scattering only their own synchrotron photons 
(Synchrotron Self-Compton, SSC for short), or also scattering radiation produced externally to the jet 
(i.e. External Compton, EC for short).
The latter process is likely to be important especially in those sources for which the second component is 
largely dominating the SED.
This usually occurs for flat spectrum radio quasars (FSRQ), which are among the most luminous persistent 
sources of the Universe.
As such, they can be detected in almost all bands also at high redshifts, 
well above $z>5$ (Romani et al. 2004, Sbarrato et al. 2012).

The most powerful FSRQ are therefore very bright hard X-ray and $\gamma$-ray sources, with the hard X-ray 
band (20--150 keV) more effective in selecting bright FSRQ at $z>4$ (Ajello et al. 2009, Ghisellini et al. 2010a), 
while the $\gamma$-ray band (0.1--10 GeV) is very effective up to $z=2-3$ (Ghisellini et al. 2011, 
Shaw et al. 2013, Giommi et al. 2013). 
Hard X-ray and $\gamma$-ray surveys recently obtained with the Burst Alert Telescope 
(BAT, onboard the {\it Swift} satellite, Gehrels et al. 2004) 
and the Large Area Telescope (LAT, on board {\it Fermi}, Atwood et al. 2009) instruments 
(Ajello et al. 2009, Cusumano et al. 2010, Nolan et al. 2012, Ajello et al. 2014) provide us 
the opportunity to see if the jet properties evolve with cosmic time 
(Volonteri et al. 2011; Ghisellini et al. 2011, 2013). 
Moreover, the nonÐ-thermal SED of these powerful sources leaves the UV band unhidden (the synchrotron hump peaks at much 
smaller frequencies, and the high-energy hump at much larger energies; Sbarrato et al. 2013), and at these frequencies
the thermal accretion flux can emerge and be detected.
This allows us to study the relation between the jet power and the accretion luminosity.

It is clear from the above that the hard X-ray and $\gamma$-ray bands are crucial to characterize the 
properties of these sources. 
In particular, the hard X-ray band (10--100 keV) is where the Compton component is rapidly rising, 
but it is also the band in which up to now we did not have very sensitive instruments (e.g. compared to those 
at softer X-rays). 
Thanks to the advent of the {\it Nuclear Spectroscopic Telescope Array} ({\it NuSTAR}) satellite (Harrison et al. 2013),
it is now possible to obtain detailed X-ray spectra in the 3--79 keV band for moderately bright X-ray sources. 
Therefore we selected two bright blazars,  PKS~2149--306
at redshift $z$=2.345 and S5~0836+710 at $z$=2.172,  and organised simultaneous observing campaign with the 
{\it NuSTAR} and {\it Swift} satellites plus on-ground optical/NIR observations for PKS~2149--306.
Both are detected in the $\gamma$-ray band by {\it Fermi}--LAT.

Both sources are high-redshift, bright blazars with a SED dominated by the Compton component. 
PKS~2149--306 was detected up to 100 keV with the 
PDS detector on board the {\it Beppo}SAX satellite with a very flat hard X-ray spectrum 
(energy spectral energy index $\alpha \simeq 0.4$, Elvis et al. 2000). 
A flat hard X-ray spectrum up to 100 keV was later also detected with {\it INTEGRAL} 
(Bianchin et al. 2009) and {\it Swift}/BAT (Sambruna et al. 2007).
This source is detected in the $\gamma$-ray band by the {\it Fermi}--LAT detector (Nolan et al. 2012).
S5~0836+710 has been detected up to 100 keV with the PDS detector on board the {\it Beppo}SAX satellite 
with a very flat hard X-ray spectrum (energy index $\alpha  \simeq 0.4$, Tavecchio et al. 2000). 
The source has also been detected both by {\it INTEGRAL} (Beckmann et al. 2006) 
and {\it Swift}/BAT (Sambruna et al. 2007).
In particular, the BAT detection shows a steeper spectrum, $\alpha \simeq 0.8$, and a flux a factor of 
$\sim 5$ weaker than the one recorded with  {\it Beppo}SAX (Sambruna et al. 2007).
This blazar is a bright and variable $\gamma$-ray source already detected by the Energetic 
Gamma-ray Experiment Telescope (EGRET, onboard the {\it Compton Gamma Ray Observatory}, Hartman et al. 1999), 
with flare activity recently seen with the {\it Fermi}--LAT detector (Akyuz et al. 2013).

Here we present the results of simultaneous observations obtained in the X-ray band with {\it NuSTAR} and
with the X-ray Telescope (XRT, Burrows et al. 2005) on board {\it Swift} for the two blazars. 
For both sources we repeated these observations over a time scale of a few months to check for source variability.
The X-ray data are complemented with simultaneous optical/UV data taken with the 
Ultra-Violet Optical Telescope (UVOT, Roming et al. 2005)  on board {\it Swift}
and, for PKS~2149--306, with optical/NIR data from the robotic 
Rapid Eye Mount (REM) telescope (Zerbi et al. 2004) located in La Silla, Chile.
For both sources we also analyze the {\it Fermi}--LAT data over a 1 year time scale centered on 
our {\it NuSTAR} and {\it Swift} observations. While we were in the process of submitting this work,
a paper appeared on astro-ph presenting these data of S5~0836+710 (Paliya 2015). The paper of
Paliya is more concentrated on the $\gamma$-ray variability, but it also analyses and discusses  the SED
of this source. We will briefly compare our results with the one of that paper in the discussion.

In this work, we adopt a flat cosmology with $H_0=70$ km s$^{-1}$ Mpc$^{-1}$ and
$\Omega_{\rm M}=0.3$.

\section{Observations and Data Analysis}
\label{sec-data}

The {\it NuSTAR}, {\it Swift} (and for PKS~2149--306 also REM)
observations are integrated with data from the {\it Wide--field Infrared Survey Explorer} 
({\it WISE}\footnote{Data retrieved from the {\it WISE} All--Sky Source Catalog: 
\url{http://irsa.ipac.caltech.edu/.}}) 
satellite (Wright et al.\ 2010) and with archival data from NASA/IPAC Extragalactic 
Database (NED) and the ASI Science Data Center (ASDC).

\subsection{{\it NuSTAR} observations}

The {\it NuSTAR} satellite carries two co-aligned hard X-ray telescopes, 
each consisting of a mirror module focusing 
high-energy X-ray photons in the band $3-79$ keV onto two independent shielded focal
plane modules (FPMs), referred to here as FPMA and FPMB (Harrison et al. 2013).

The {\it NuSTAR} satellite observed PKS 2149--306 on 2013 December 17
(obsID 60001099002) and on 2014 April 18 (obsID 60001099004). 
The total net exposure times were 38.5 ks and 44.1 ks, respectively.
S5 0836+710 was observed by {\it NuSTAR} on 2013 December 15
(obsID 60002045002) and on 2014 January 18 (obsID 60002045004), for
total net exposure times of 29.7 ks and 36.4 ks, respectively.

The FPMA and FPMB data sets were first processed with the NuSTARDAS
software package (v.1.4.1) jointly developed by the ASI Science Data
Center (ASDC, Italy) and the California Institute of Technology
(Caltech, USA). Event files were calibrated and cleaned with standard
filtering criteria with the {\it nupipeline} task using the version
20140414 of the {\it NuSTAR} CALDB.

For all four observations the FPMA and FPMB spectra of the sources were
extracted from the cleaned event files using a circle of 20 pixel
($\sim 49$ $\arcsec$) radius, while the background was extracted from two
distinct nearby circular regions of 50 pixel radius. 
The ancillary response files were generated with the {\it numkarf} task, applying
corrections for the point spread function (PSF) losses, exposure maps and vignetting. All
spectra were binned to ensure a minimum of 20 counts per bin.
Both sources are quite bright and well detected by {\it NuSTAR} up to 79 keV.

\subsection{{\it Swift} observations}

{\it Swift} (Gehrels et al. 2004) observed PKS 2149--306 on 2013 December 16--17
(obsIDs 00031404013 and 00031404014) and on 2014 April 18 (obsID
00031404014) and S5 0836+710 was observed on 2013 December 16 (obsIDs
00080399001) and on 2014 January 18 (obsID 00080399002).

\subsubsection{XRT observations}

The XRT on board {\it Swift} is sensitive
to the $0.3-10$ keV X-ray energy band (Burrows et al. 2004).
All XRT observations were carried out using the most sensitive Photon
Counting (PC) readout mode. 
The XRT data sets were first processed with the XRTDAS software package 
(v.3.0.0) developed at the ASI Science Data Center (ASDC) and distributed by 
HEASARC within the HEASoft package (v. 6.16). 
Event files were calibrated and cleaned with standard filtering criteria 
with the {\it xrtpipeline} task using the calibration files available in 
the version 20140709 of the {\it Swift}--XRT CALDB.

The two individual XRT event files for the 2013 December observations of PKS~2149--306
were merged together using the XSELECT package for a total net exposure time of 8.0 ks.  
The net exposure time of the 2014 April observation was 6.4 ks. 
For the two observations of S5 0836+710 the total net exposure times were 1.9 ks
and 4.7 ks, respectively. 
For all observations the energy spectra were extracted from the summed
cleaned event files. 
Events for the spectral analysis were selected within a circle of 20 pixel 
($\sim 47$ $\arcsec$) radius, which encloses about 90\% of the PSF,
centered on the source position. 
The background was extracted from a nearby circular region of 50 pixel radius. 
The ancillary response files were generated with the {\it xrtmkarf} task, applying corrections
for the PSF losses and CCD defects using the cumulative exposure map.
The source spectra were binned to ensure a minimum of 20 counts per bin.

\subsubsection{UVOT observations}

UVOT observations were performed with all six optical and UV
lenticular filters (namely $W2$, $M2$, $W1$, $u$, $b$, $v$) (Roming et al. 2005). 
We performed aperture
photometry for all filters in all observations using the standard
UVOT software distributed within the HEAsoft package (version 6.15.1)
and the calibration included in the latest release of the CALDB. 
Counts were extracted from aperture of 5 arcsec radius for all filters 
and converted to fluxes using the standard zero points (Poole et al. 2008). 
The fluxes were then de-reddened using the appropriate
values of $E(B-V)$ taken from Schlegel et al. (1998) and Schlafly et
al. (2011) with $A_{\lambda}/E(B-V)$ ratios calculated for UVOT filters
using the mean Galactic interstellar extinction curve from Fitzpatrick (1999). 
These fluxes were then used to build the SEDs of both sources (see Fig. \ref{sed}). 
No variability was detected within single exposures in any filter. 
The processing results were carefully verified checking for
possible contamination from nearby objects within the source apertures and
from objects falling within background apertures. 
Both sources are well detected in all filters.
The values of the magnitudes in the Vega system are given in Table \ref{uvot}.

\begin{table*} 
\centering
\centerline{PKS~2149--306}
\begin{tabular}{lllllll}
\hline
\hline
Date & $v$ & $b$ & $u$ & $W1$ & $M2$ & $W2$ \\
\hline   
17 Dec 2013    & $17.83 \pm 0.11$ & $17.86 \pm 0.06$ & $17.14 \pm 0.06$ & $18.22 \pm 0.10$ & $20.46 \pm 0.07$ &  ... \\  
18 Apr 2014    & $17.61 \pm 0.05$ & $17.75 \pm 0.02$ & $17.14 \pm 0.02$ & $18.06 \pm 0.04$ & $20.00 \pm 0.11$ & $20.12 \pm 0.08$ \\   
\hline
\hline 
\end{tabular}
\vskip 0.2 truecm
\centerline{S5 0836+710}
\vskip 0.1 truecm
\begin{tabular}{lllllll}
\hline
\hline
15 Dec 2013    &...  &... &... &...  &$17.45 \pm 0.03$ &... \\  
18 Jan 2014    & $16.95 \pm 0.03$ & $17.12 \pm 0.02$ & $16.22 \pm 0.02$ & $16.92 \pm 0.03$ & $17.53 \pm 0.04$ & $17.91 \pm 0.03$ \\   
\hline
\hline 
\end{tabular}

\vskip 0.4 true cm
\caption{UVOT Vega $v$, $b$, $u$, $W1$, $M2$, $W2$ observed magnitudes of PKS~2149--306 and S5~0836+710
(magnitudes not corrected for Galactic extinction). 
}
\label{uvot}
\end{table*}

\subsection{REM observations}

We observed PKS~2149--306 with the Rapid Eye Mount Telescope (REM, Zerbi et al. 2004), a
robotic telescope located at La Silla Observatory
(Chile).
It performed photometric observations in the optical {\it gri} and near-infrared (NIR) 
$JHK$ filters in the nights of 2013 December 15 and 19. 
REM has a Ritchey--Chretien 
configuration with a 60 cm f/2.2 primary and an overall f/8 focal ratio in a fast-moving 
alt-azimuth mount that provides two stable Nasmyth focal stations. 
The two cameras, REMIR (Conconi et al. 2004) for the NIR and ROS2 (Molinari et al. 2014) 
for the optical, both have the same field of view of $10 \times 10$ $\arcmin$. 
The telescope is able to operate in a fully autonomous way (Covino et al. 2004) 
and data are reduced and analyzed following standard procedures. 
Aperture photometry was derived by means of custom tools\footnote{https://pypi.python.org/pypi/SRPAstro.FITS/} 
and calibration was based on objects in the field of view reported in 
the APASS\footnote{http://www.aavso.org/apass} and 2MASS\footnote{http://www.ipac.caltech.edu/2mass/} 
catalogues in the optical and NIR, respectively.

Table \ref{rem} reports the observed {\it gri} (AB) and $JHK$ (Vega) photometry measured on the two nights, 
not corrected for the Galactic extinction of $E(B-V)=0.02$ from Schlegel et al. (1998).
 
\begin{table*} 
\centering
\begin{tabular}{lllllll}
\hline
\hline
Date & $g$ &$r$ &$i$ &$J$ &$H$ &$K_{s}$ \\
\hline   
15 Dec 2013    & $17.82 \pm 0.05$ & $17.58 \pm 0.06$ & $17.40 \pm 0.06$ & $16.56 \pm 0.06$ & $16.00 \pm 0.08$ & $15.10 \pm 0.11$ \\  
19 Dec 2013    & $17.75 \pm 0.05$ & $17.36 \pm 0.06$ & $17.38 \pm 0.06$ & $16.58 \pm 0.07$ & $15.89 \pm 0.07$ & $15.22 \pm 0.13$ \\   
\hline
\hline 
\end{tabular}
\vskip 0.4 true cm
\caption{REM AB ({\it gri}) and VEGA ($JHK$) observed magnitudes of PKS~2149--306  
(magnitudes not corrected for Galactic extinction). 
}
\label{rem}
\end{table*}

\subsection{{\it Fermi}--LAT observations}

Both sources are also bright $\gamma$-ray emitters and are regularly detected by {\it Fermi}--LAT.
We analyzed the data collected for one year between 2013 June 01 -- 2014 June 01 (MJD 56444 -- 56809)
following the standard procedure\footnote{http://fermi.gsfc.nasa.gov/ssc/data/analysis/}, 
using the {\it Fermi}--LAT analysis software \texttt{ScienceTools} \texttt{v9r34p1} with the 
\texttt{P7REP\_SOURCE\_V15} instrument response functions. Events in the energy range
100 MeV--300 GeV were extracted within a $15^{\circ}$ acceptance cone of the Region of     
Interest (ROI) centered on the location of each source. Gamma-ray fluxes and spectra were
determined by an unbinned maximum likelihood fit with \texttt{gtlike}. The background model
included all known $\gamma$-ray sources within the ROI from the 2nd {\it Fermi}--LAT catalog 
(Nolan et al. 2012).\footnote{ In order to test the effect of possible new gamma-ray sources 
not included in 2FGL, we reanalyzed the data using the Fermi-LAT 4-year point source catalog
(3FGL, Acero et al. 2015), finding fully consistent results.}
Additionally, the model included the isotropic and Galactic diffuse emission components.
Flux normalization for the diffuse and background sources were left free in the fitting procedure.

\begin{figure*}
\vskip -0.5 true cm 
\epsscale{1.7}
\plotone{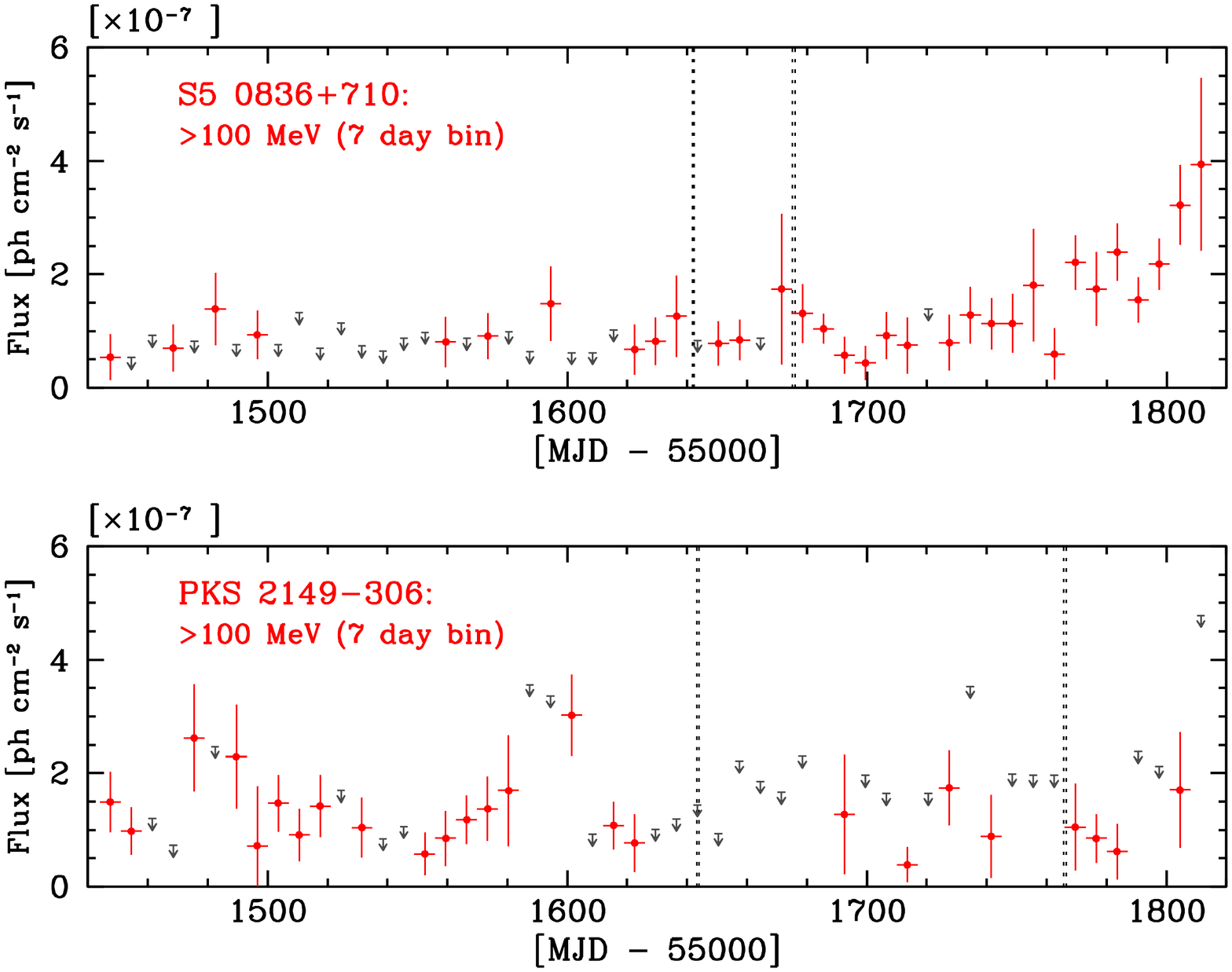} 
\vskip -0.5 true cm
\caption{
Gamma-ray light curves ($>100$ MeV) of S5~0836+710 (top panel) and PKS~2149--306 (bottom panel) 
as observed by {\it Fermi}/LAT. 
The vertical dotted lines mark the dates of the {\it NuSTAR} observations.
Upper limits correspond to TS $<$ 4 (see text).
} 
\label{LAT}
\end{figure*}

The {\it Fermi}--LAT light-curves above 100 MeV binned over a time scale of seven days are shown 
in Fig. \ref{LAT} for both sources. 
We mark with vertical dotted lines the dates of the {\it NuSTAR} observations.
For bins with test statistic TS $<$ 4, 95\% confidence level upper limits are plotted
(for the meaning of TS see Mattox et al. 1996).
We fitted the 1--year {\it Fermi}--LAT spectrum for each source with a power--law model, finding a 
best-fit photon spectral index of  $\Gamma=2.89 \pm 0.09$ (TS=210) and $\Gamma=2.69 \pm 0.10$ (TS=284) 
for PKS~2149--306 and S5~0836+710, respectively (errors are at 68\% confidence interval 
for one parameter of interest). 
The corresponding average 1--year fluxes are $F(>100$ MeV)$=(8.4\pm0.9) \times 10^{-8}$  ph cm$^{-2}$ s$^{-1}$
and $F(>100$ MeV)$=(5.6\pm0.8) \times 10^{-8}$ ph cm$^{-2}$ s$^{-1}$.

In the case of S5~0836+710 we have enough statistics to extract a 1--month 
(2014 January 1 -- 2014  February 1) source spectrum centered around the second {\it NuSTAR} observation. 
The fitting result yielded $\Gamma=2.7 \pm 0.2$ (TS=46) with an average  flux above 
100 MeV of $(8.7\pm1.9) \times 10^{-8}$ ph cm$^{-2}$ s$^{-1}$. From Fig. \ref{LAT} it is also apparent that a couple of months later the source became 
much more active in the {\it Fermi}--LAT band. 

\begin{table*} 
\centering
\centerline{PKS2149--306}
\vskip 0.1 truecm
\begin{tabular}{lllllll}
\hline
\hline
 Date &  $\Gamma_1$ & $\Gamma_2$ & break & $F_{\rm 2-10\, kev}$  & $F_{\rm 10-40/, kev}$  & $\chi^2$ / dof \\
& & &keV &$\ergcms$ &$\ergcms$ & \\
\hline  
 \hline
2013 Dec 17  & $0.94_{-0.08}^{+0.07}$ & $1.35_{-0.01}^{+0.02}$ & $2.61_{-0.46}^{+0.67}$  & $2.2\times10^{-11}$ & $5.0\times10^{-11}$ & 1095.8 / 1108 \\
2014 Apr 18 & $0.97_{-0.09}^{+0.07}$ & $1.46_{-0.02}^{+0.02}$ & $3.25_{-0.63}^{+1.36}$  & $1.9\times10^{-11}$ & $3.7\times10^{-11}$ & 936.2 / 952 \\
\hline
\hline 
\end{tabular}
\vskip 0.2 truecm
\centerline{S5 0836+710}
\vskip 0.1 truecm
\begin{tabular}{lllllll}
\hline
\hline
  Date &  $\Gamma_1$ & $\Gamma_2$ & break & $F_{\rm 2-10\, kev}$  & $F_{\rm 10-40\, kev}$  & $\chi^2$ / dof \\
 & & & keV &  $\ergcms$ & $\ergcms$ & \\
\hline  
\hline
2013 Dec 15 & $1.03_{-0.32}^{+0.20}$ & $1.66_{-0.02}^{+0.02}$ & $1.73_{-0.48}^{+1.27}$  & $1.6\times10^{-11}$ & $2.3\times10^{-11}$ & 642.6 / 611\\
2014 Jan 18 & $1.18_{-0.10}^{+0.08}$ & $1.66_{-0.01}^{+0.02}$ & $2.84_{-0.62}^{+1.03}$  & $2.5\times10^{-11}$ & $3.6\times10^{-11}$ & 896.4 / 892 \\
\hline
\hline 
\end{tabular}
\caption{Parameters of the X-ray spectral analysis for the simultaneous fit 
of the {\it NuSTAR} and {\it Swift}/XRT data. The errors are at 90\% level 
of confidence for one parameter of interest. 
Fluxes are corrected for the galactic absorption.
}
\label{xspec}
\end{table*}

\begin{figure*}
\vskip 0 true cm 
\epsscale{1}
\plotone{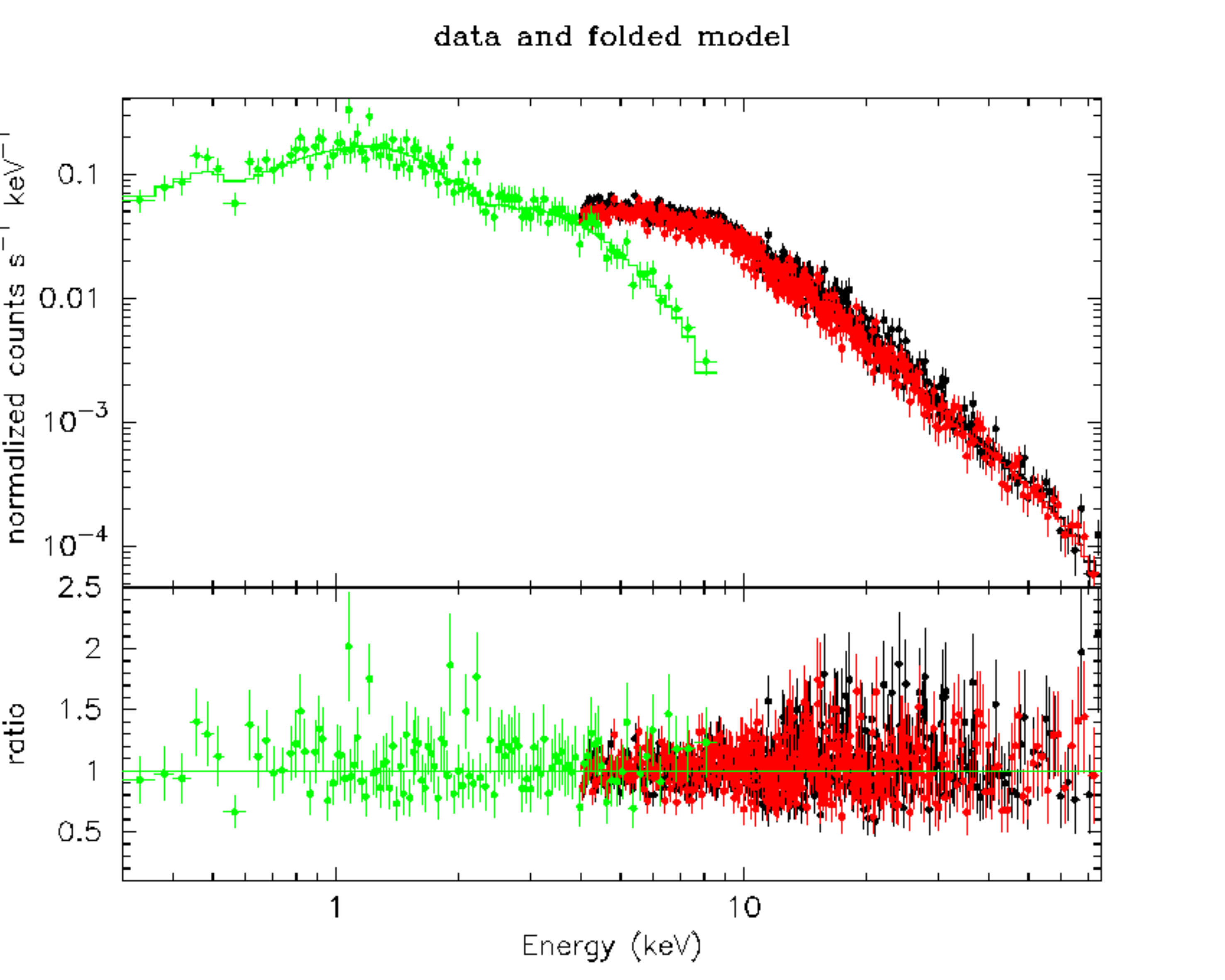}
\vskip 0 true cm
\caption{
X-ray spectrum of PKS~2149--306 as observed by {\it Swift}/XRT and {\it NuSTAR} on 2014 April 18, 
together with the broken power-law best-fit model. 
{\it NuSTAR} data are filled circles and triangles (black and red, in the electronic version), 
while {\it Swift}/XRT data are empty squares (green in the electronic version).
} 
\label{spettrox}
\end{figure*}

\subsection{X-ray spectral analysis}
\label{xray_an}

For all observations of both sources simultaneous fits of the XRT 
and {\it NuSTAR} spectra were performed using the XSPEC package. 
In both cases a broken power-law model with an absorption hydrogen-equivalent column density 
fixed to the Galactic value was adopted ($N_{\rm H}=1.6 \times 10^{20}$ cm$^{-2}$ and 
$N_{\rm H}=2.8 \times 10^{20}$ cm$^{-2}$ for PKS~2149--306 and S5~0836+710, respectively, Kalberla et al. 2005) . 
To allow for the cross-calibration uncertainties between the three 
telescopes (two {\it NuSTAR} and one {\it Swift}), a multiplicative constant factor 
has been included, kept equal to 1 for the FPMA spectra and free to vary for the FPMB and XRT spectra. 
In the case of FPMB the difference is in the range of 2--4\%,
while for XRT it is somewhat larger but alway less than 10\%.
This is consistent with the values usually found for other sources.
We find that this model provides a good description of the observed spectra in the 0.3--79 keV energy band. 
The results of the spectral fits are in Table \ref{xspec}, while Figure \ref{spettrox} 
shows an example of the XRT and {\it NuSTAR} spectra together with the best-fit model 
for the observation of PKS~2149--306 on 2014 April 18.
To test the robustness of the spectral curvature, we estimated the instrumental cross-calibration 
factors by fitting the data in a common energy band (3-9 keV) adopting a power-law model. 
We found fully consistent best-fit values of the cross-calibration factors. We then verified that 
the broken power-law best-fit spectral results were unchanged using these values.
Besides using a broken power-law model we tried a simpler power-law model, but this did not provide
a good fit to the data for either source. 
In fact, in all cases the reduced $\chi^2_r$ was larger than 1.1--1.2 with more than 900 d.o.f.. 
The F-test shows that the improvement of these values obtained with the broken
power-law model is significant with a probability smaller than $10^{-30}$ in three cases and 
smaller than $10^{-11}$ in one case.
Moreover, when using the power-law model the constant value included to take into account
the cross-calibration uncertainties was too large or too low (e.g. 20--30\% difference) with respect to
the values usually found, another indication that the  single power-law model is not correct. 

\begin{figure*}
\vskip 0 true cm 
\epsscale{1.}
\plotone{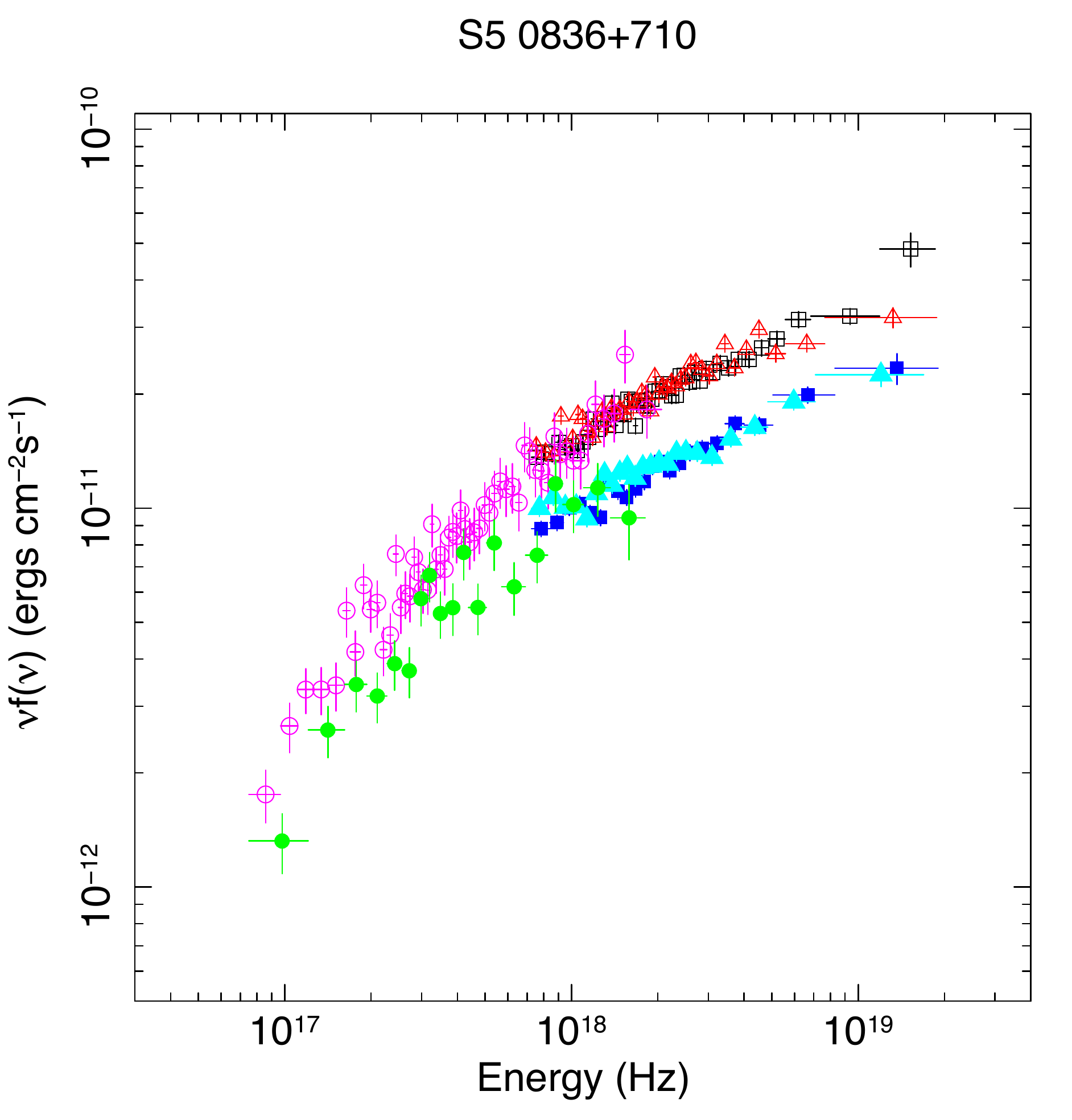}
\plotone{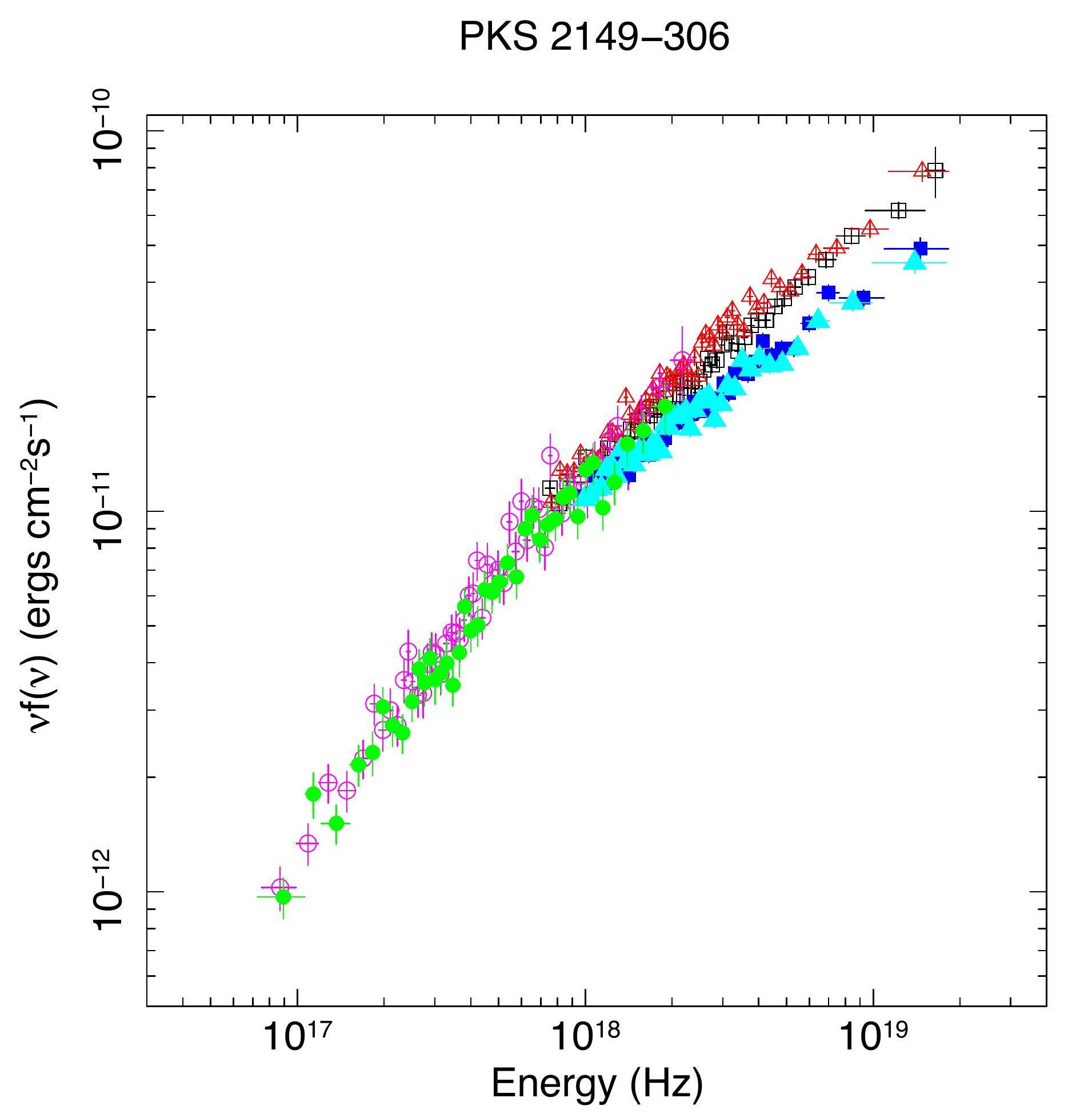}
\vskip 0 true cm
\caption{
X-ray SED of S5~0836+710 (left panel) and of PKS~2149--306 
(right panel) as observed by {\it Swift}/XRT and {\it NuSTAR} on four different 
dates (see text). 
XRT data points are filled--empty circles (green and purple in the electronic version), 
{\it NuSTAR}/FPMA data points are filled-empty squares (red and blue in the electronic version), 
{\it NuSTAR}/FPMB data points are filled-empty triangles (black and magenta in the electronic version).
}
\label{SEDX}
\end{figure*}

In Table \ref{xspec} we also provide the fluxes in the two bands $2-10$ and $10-40$ 
keV for the four observations. 
Both sources have varied between the two observations, spanning a time scale 
of one month for S5~0836+710 and of four months for PKS~2149--306. 
No fast variability is observed during the single observations.
However, while for S5~0836+710 there is an increase by a factor of $\sim 1.5$ over the full 
0.3--79 keV band, with a hint of lower variability below $\sim 1$ keV, for PKS~2149--306 a 
variability of $\sim 30\%$ is present essentially only above 10 keV. 
To better show this we plot all XRT and {\it NuSTAR} spectra in Figure \ref{SEDX}.

\section{Overall Spectral Energy Distribution}

\begin{figure*}
\epsscale{1.}
\plotone{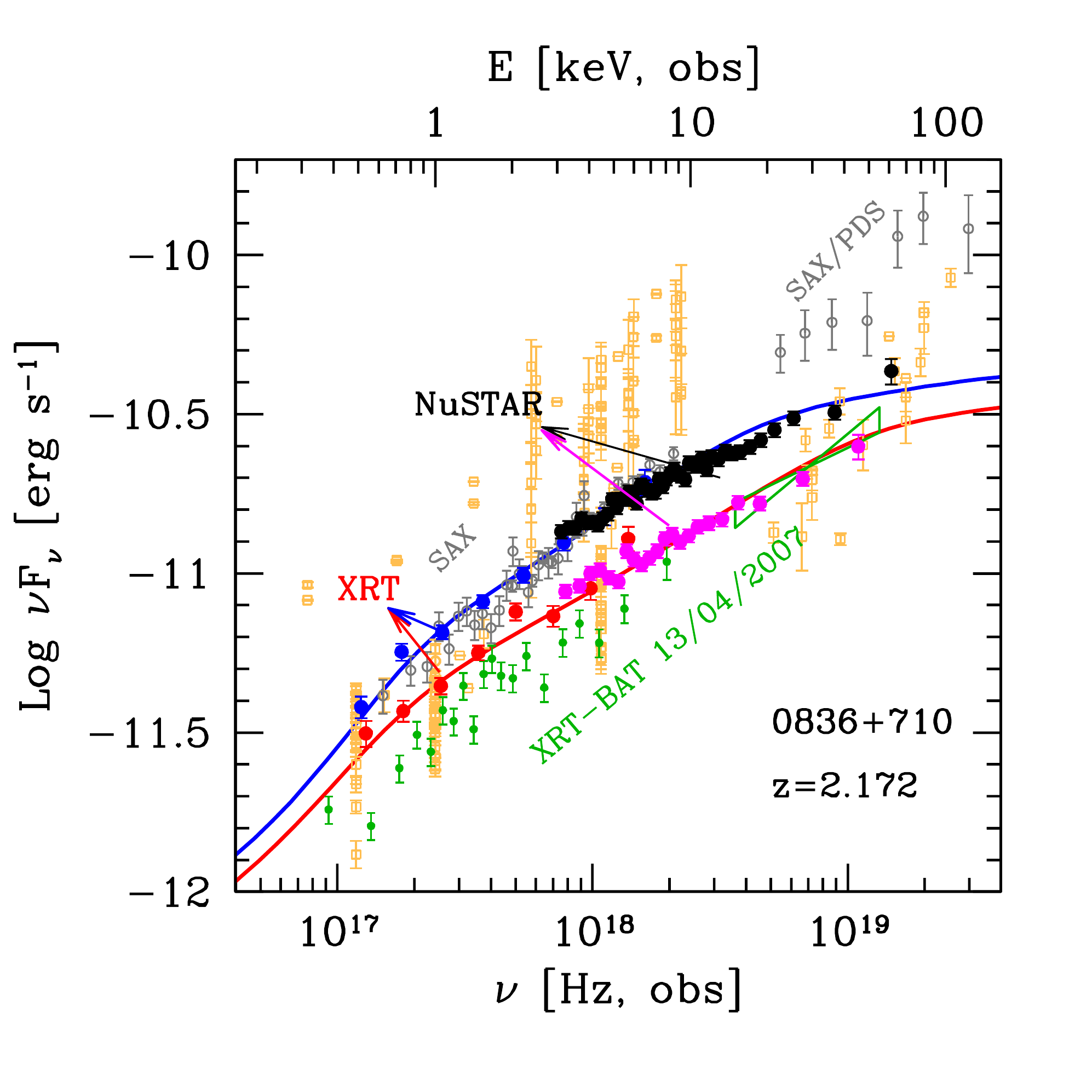}
\plotone{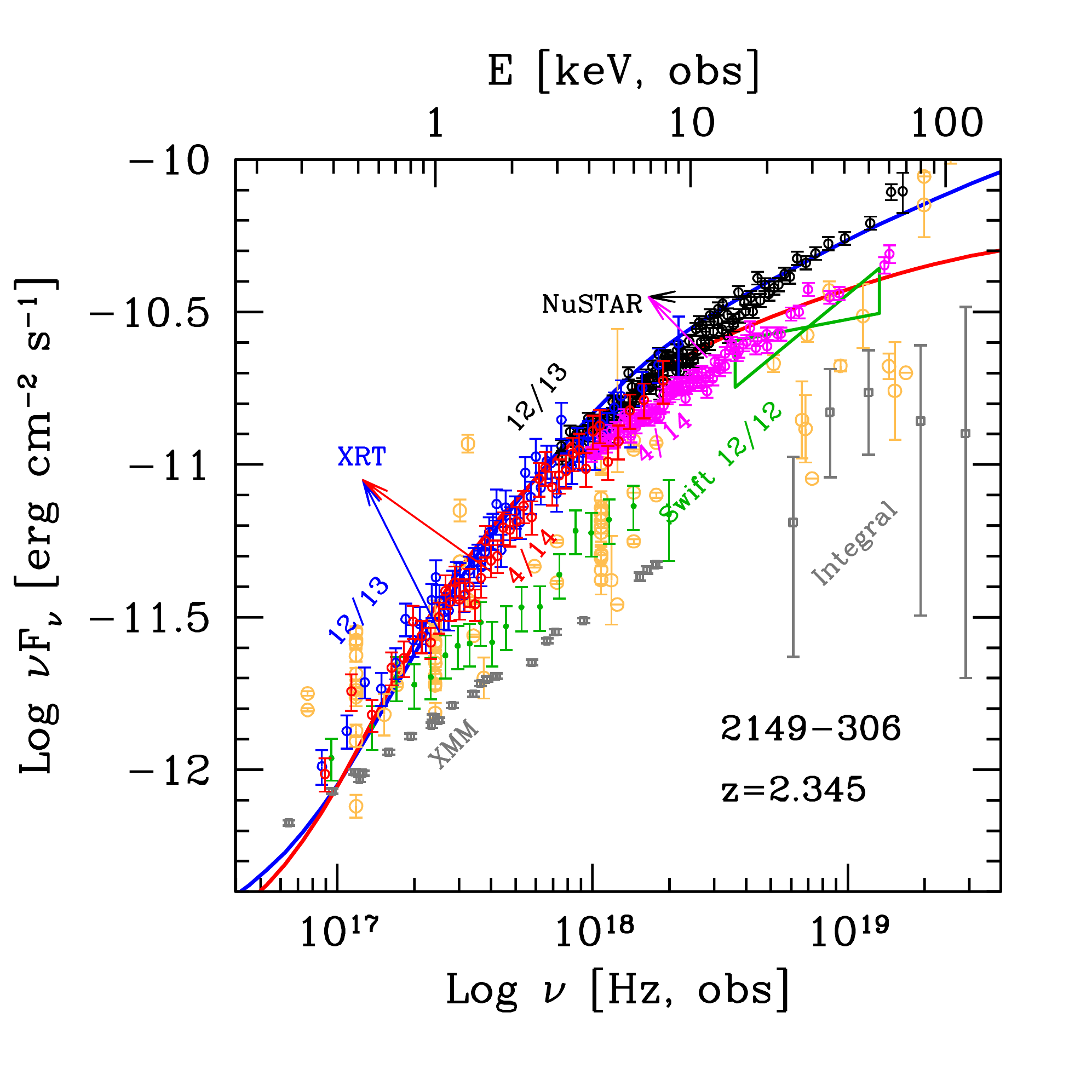}
\vskip -0.5 true cm
\caption{
X-ray SED of S5~0836+710 (left panel) and PKS 2149--306 (right panel) as observed 
by {\it Swift}/XRT and {\it NuSTAR}, compared with previous observations
by {\it Swift} (green dots), {\it INTEGRAL}, {\it XMM} and {\it Beppo}SAX (grey),
and other archival observations (orange).
Our XRT and {\it NuSTAR} data points for the two observing periods are labelled.
The solid lines refer to the model used to explain the overall SED, as shown in Fig. \ref{sed}.
} 
\label{zoomx}
\end{figure*}

Fig. \ref{zoomx} shows the X-ray data of both sources, and
compares them with archival observations.
For S5 0836+710 the two observations correspond to different
flux levels, and slightly different spectra.
The XRT spectrum  smoothly joins the {\it NuSTAR} data points
in both observations.
The two data sets differ mainly around 10 keV, by a factor $\sim$1.6,
with less variations at both ends of the observed X-ray band.
On the contrary, for PKS 2149--306 the variability amplitude 
monotonically increases at higher frequencies, while the two spectra
are very similar in the soft X-ray band.
The archival data (orange symbols) indicate that both sources
(but especially S5 0836+710) can vary in X-rays by about an order of magnitude.
Note also the much softer archival spectrum of PKS 2149--306 by {\it XMM} (and by {\it INTEGRAL}).

\begin{figure*}
\vskip -0.5 true cm 
\epsscale{1.}
\plotone{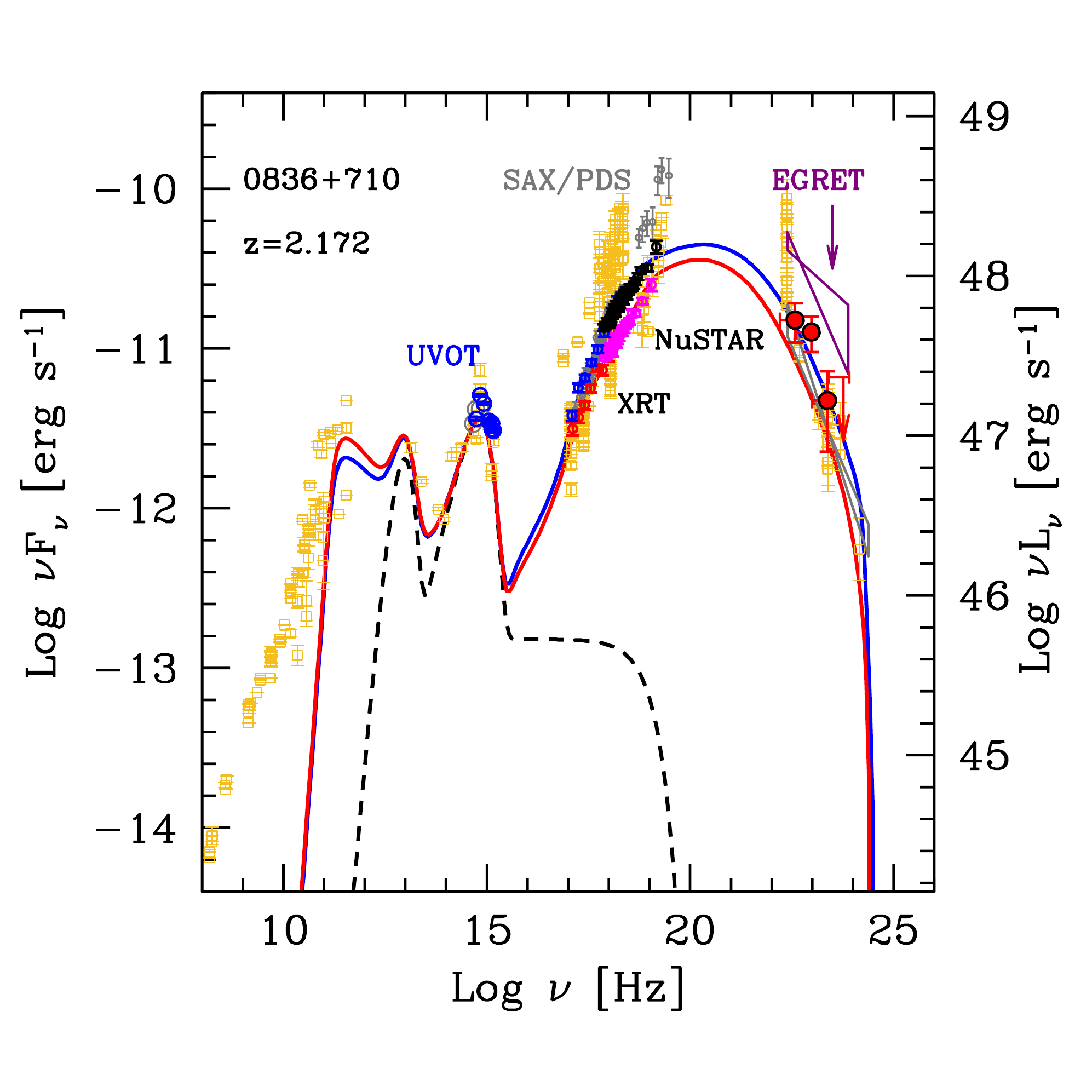}
\plotone{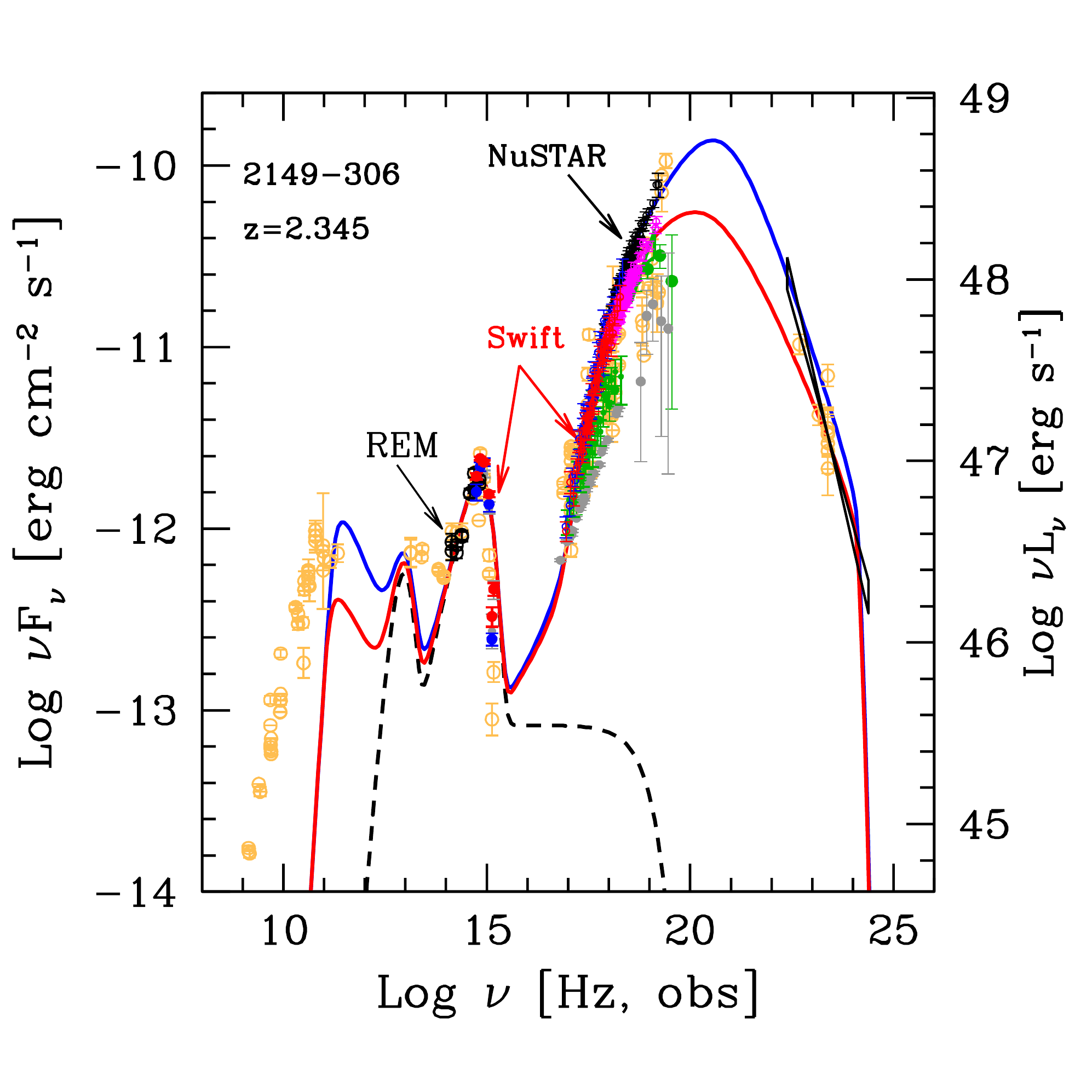}
\vskip -0.5 true cm
\caption{
Overall SED of S5 0836+710 (left panel) and of PKS 2149--306 (right panel).
The indicated REM, UVOT, XRT and {\it NuSTAR} data are simultaneous.
For S5 0836+710, the {\it Fermi}--LAT $\gamma$-ray data (red points encircled by black circles)
are time integrated over a month (2014 Jan. 1st--2014 Feb. 1st), centered on the second  
{\it NuSTAR} observation, 
while the bow-tie represents the result of the one-year integration (2013 June 1st--2014 June 1st). 
Fot PKS 2149--306, the {\it Fermi}--LAT data (bow tie) are the result of one-year integration
(2013 June 1st--2014 June 1st). 
The solid lines refer to the one-zone leptonic model discussed in the text. The dashed
black line is the contribution of the accretion disk, the IR torus, and the X-ray corona.
We also show archival data (empty circles, orange in the electronic version).
} 
\label{sed}
\end{figure*}

Fig. \ref{sed} shows the overall SED of the two sources.
Both of them show, besides the typical double-peak SED of blazars, 
a third narrow peak in the IR--UV band, which we identify as thermal emission 
from the accretion disk.
Fitting it with a standard Shakura \& Sunyaev (1973) disk we find both the disk 
luminosity $L_{\rm d}$ and the black hole mass $M$ (as listed in Table \ref{para} 
and Table \ref{powers}, 
see Calderone et al. 2013 for a full discussion about this method).
For a given efficiency $\eta$ (defined through $L_{\rm d}=\eta \dot M c^2$),
and when the peak of the disk emission is visible, this method
returns a value of the black hole mass with a relatively
small uncertainty (factor $\sim$1.5), better than the
virial method (factor 3--4, Vestergaard et al. 2006; Park et al. 2012).
Adopting $\eta=0.08$,
we find a black hole mass of $5\times 10^9 M_\odot$ 
for S5 0836+710) and  $3.5\times 10^9 M_\odot$ for 
PKS 2149--306.\footnote{The values of the black hole masses are somewhat 
dependent upon $\eta$: the larger $\eta$, 
the greater the black hole mass.
For instance, adopting $\eta=0.06$ (0.15), we would obtain 
$M=3\times 10^9 M_\odot$ ($6\times 10^9 M_\odot$) for S5 0836+710.
}
These values differ slightly from what we reported previously using the same
method (in Ghisellini et al. 2010a, hereafter GG10). 
This is due to the better coverage of the IR part of the spectrum 
now available from {\it WISE} and REM.
With these values of the black hole mass, the disks of both sources emit 
at $\sim$20--30\% of the Eddington luminosity.
The blazar S5 0836+710 has also been monitored for reverberation mapping
by Kaspi et al. (2007).
Using the observed lag between the continuum, the C\,{\sc iv} FWHM 
and Eq. 5 of Kaspi et al. (2000), they obtained a black hole mass $M=2.6\times 10^9 M_\odot$.
Instead, using the UV luminosity, the CIV FWHM and Eq. 7 of Vestergaard \& Peterson (2006),
they derive $M=1.8\times 10^{10} M_\odot$.

As discussed also in GG10, these sources are {\it weak} in the $\gamma$-ray band.
This is due to two effects.
The first is a k-correction effect: for increasing redshifts, the 
observed hard X-ray spectrum is closer to the high-energy peak, and thus brighter.
The second effect is due to the intrinsic shift of the high-energy peak frequency
as the bolometric non-thermal luminosity is increased.
This implies that the most powerful jetted sources can be better found
in hard X-ray surveys rather than in $\gamma$-ray surveys (GG10).
For S5 0836+710 {\it Fermi}--LAT detected the source in one-month integration times
around the second {\it NuSTAR} observation (filled red points with black circles)
while PKS 2149--306 is detected using one-year integration times (from June 1st 2013 to June 1st 2014).
Both $\gamma$-ray spectra are steep, as usual for powerful FSRQ (see e.g. Nolan et al. 2012). 
The fact that the synchrotron far--mm--IR spectrum is steep is consistent with 
the assumption that both the low and high-energy non-thermal peaks of the SED 
are produced by the same population of electrons.

We applied a one-zone leptonic model, fully described in Ghisellini \& Tavecchio (2009),
to interpret the SED of the two sources, and to explore the
variability that both sources experienced between the two {\it Swift+NuSTAR} observations.
In brief, the model assumes that the emitting source is a homogeneous sphere located at a 
distance $R_{\rm diss}$ from the black hole, in a conical jet of semi-aperture angle $\phi =0.1$ rad. 
The BLR is assumed to be a spherical shell of radius $R_{\rm BLR}=10^{17} L_{\rm d,45}^{1/2}$ cm
and the size of the IR torus is assumed to be $R_{\rm torus}=2\times 10^{18} L_{\rm d, 45}^{1/2}$ cm
(here $L_{\rm d, 45}$ is the disk luminosity in units of $10^{45}$ erg s$^{-1}$).
The source is assumed to move with a bulk Lorentz factor $\Gamma$ in a direction making an angle 
$\theta_{\rm v}$ with the line of sight.
The magnetic field $B$ (as measured in the comoving frame) depends upon the distance
from the black hole and the bulk Lorentz factor, to yield a constant
Poynting flux $P_{\rm B} \propto B^2R^2_{\rm diss}\Gamma^2$.

The energy distribution of the emitting particles is found through the continuity equation,
accounting for continuous injection, radiative cooling, and electron-positron pair production.
Electrons (with random Lorentz factors $1<\gamma<\gamma_{\rm max}$)
are assumed to be injected with a total power $P^\prime_{\rm i}$
(in the comoving frame) throughout the source with a broken power-law distribution, 
$\propto \gamma^{-s_1}$ and $\propto \gamma^{-s_2}$ below and above $\gamma_{\rm b}$, respectively.
The radiative processes are synchrotron, self-Compton and external Compton with both photons from the BLR
and the torus.
We also consider the presence of an X-ray corona, emitting 30\% of $L_{\rm d}$ 
with a spectrum $F_\nu \propto \nu^{-1}\exp(-h\nu/150{\, \rm keV}$).

Due to the compactness of the source, required to account for the fast variability,
the synchrotron emission is self-absorbed at radio frequencies, up to hundreds of GHz
(observer frame).
Therefore these models cannot reproduce the radio data.

\begin{table*}
\centering
\begin{tabular}{lllll lllll llll}
\hline
\hline
Name &M &$\Gamma$  &$\theta_{\rm v}$ &$R_{\rm diss}$ &$R$ &$R_{\rm BLR}$   &$P^\prime_{\rm i}$ &$B$
&$\gamma_{\rm b}$ &$\gamma_{\rm max}$ &$s_1$  &$s_2$  &$\gamma_{\rm c}$ \\
~[1]       &[2] &[3] &[4] &[5] &[6] &[7] &[8] &[9] &[10] &[11] &[12] &[13]  \\       
\hline
0836+710 H     &5e9   &16 &3 &1950  &195 &1500 &0.11 &1.11 &250 &5e3 &1.7 &3.2  &5.7 \\
0836+710 L     &5e9   &16 &3 &2100  &210 &1500 &0.09 &1.04 &190 &4e3 &1.6 &3.2  &9.8 \\
0836+710 GG10  &3e9   &14 &3 &540   &54  &1500 &0.22 &3.28 &90  &2e3 &--1 &3.6  &2.1 \\
\hline
2149--306 H    &3.5e9 &14 &3 &1365  &137 &1212 &0.2  &1.05 &75 &4e3 &0.5 &3.3   &2.9 \\
2149--306 L    &3.5e9 &14 &3 &1365  &137 &1212 &0.1  &1.05 &50 &4e3 &1   &3.0   &2.9 \\
2149--306 GG10 &5e9   &15 &3 &1200  &120 &1224 &0.18 &1.12 &60 &3e3 &0   &3.3   &1   \\
\hline
\hline
\end{tabular}
\vskip 0.2 true cm
\caption{
Input parameters used to model the SED. ``L" and ``H"stand for low and high state,
respectively; GG10 stands for Ghisellini et al. (2010a), where the two sources were
also studied.
Col. [1]: Source name and state/observation;
Col. [2]: Black hole mass in solar mass units;
Col. [3]: bulk Lorentz factor;
Col. [4]: viewing angle (degrees);
Col. [5]: distance of the blob from the black hole in units of $10^{15}$ cm;
Col. [6]: Source size in units of $10^{15}$ cm;
Col. [7]: size of the broad line region in units of $10^{15}$ cm; 
Col. [8]: power injected in the blob calculated in the comoving frame, in units of $10^{45}$ erg s$^{-1}$;
Col. [9]: magnetic field in Gauss;
Col. [10] and [11]: minimum and maximum random Lorentz factors of the injected electrons;
Col. [12] and [13]: slopes of the injected electron distribution [$Q(\gamma)$] below
and above $\gamma_{\rm b}$;
Col. [13]: value of the minimum random Lorentz factor of the electrons cooling in $R/c$.
The spectral shape of the corona is assumed to be $\propto \nu^{-1} \exp(-h\nu/150~{\rm keV})$.
}
\label{para}
\end{table*}

\begin{table*}
\centering
\begin{tabular}{lllll ll}
\hline
\hline
Name &$\log L_{\rm d}$ &$L_{\rm d}/L_{\rm Edd}$ &$\log P_{\rm r}$ &$\log P_{\rm B}$ &$\log P_{\rm e}$ &$\log P_{\rm p}$ \\
~[1] &[2]              &[3]                     &[4]              &[5]              &[6]              &[7]  \\ 
\hline
0836+710 H     &47.4 &0.3 &46.8 &46.7 &45.9 &48.6  \\
0836+710 L     &47.4 &0.3 &46.7 &46.7 &45.9 &48.5  \\
0836+710 GG10  &47.4 &0.5 &46.6 &46.4 &45.5 &48.0  \\
\hline 
2149--306 H    &47.2 &0.3 &47.1 &46.2 &45.7 &48.3  \\
2149--306 L    &47.2 &0.3 &46.7 &46.2 &45.6 &48.3  \\
2149--306 GG10 &47.2 &0.2 &46.6 &46.2 &45.3 &48.0  \\
\hline
\hline
\end{tabular}
\vskip 0.2 true cm
\caption{
Accretion and jet powers:
Col. [1]: Source name and state/observation;
Col. [2]: Logarithm of the accretion disk luminosity (units are erg s$^{-1}$);
Col. [3]: Accretion disk luminosity in Eddington units
Col. [4]--[7] logarithm of the jet power in the form of radiation ($P_{\rm r}$), Poynting flux ($P_{\rm B}$),  
bulk motion of electrons ($P_{\rm e}$) and [7] protons ($P_{\rm p}$  (assuming one cold proton
per emitting electron). Units are erg s$^{-1}$.
These values refer to one jet only.
}
\label{powers}
\end{table*}

Fig. \ref{sed} shows the models corresponding to the two states of each source
and Table \ref{para} lists the model parameters.
It also reports the parameters used in GG10 to fit another data set.
To model the two observed states, we changed a minimum number of parameters.
Both sources have $R_{\rm torus}>R_{\rm diss}>R_{\rm BLR}$.
This choice is preferred (with respect to $R_{\rm diss}$ within the BLR)
because of the small value of the high-energy peak, requiring seed photons
of frequency smaller than the hydrogen Ly$\alpha$ photons (that dominate
the BLR emission).
In this respect, {\it NuSTAR} is crucial  because the hardness of its spectrum,
together with the extrapolation from the $\gamma$-ray energies, greatly helps
in pinpointing the peak frequency of the high-energy hump.
This is the major difference with respect to the parameters reported by GG10
for S5 0836+710, which assumed $R_{\rm diss}<R_{\rm BLR}$.
 
The observed variability of S5 0836+710 is ascribed to a small difference
in the position of the emitting region, in turn implying a small change 
in the magnetic field.
Furthermore, the injected power changes by 10\% and the break energy of 
the injected electron distribution changes by 20\%.

For PKS 2149--306 the variability can be explained by changing the injected
power by a factor of 2, and by a 50\% change in the break energy of the 
injected electron distribution. 

Comparing these changes with the distribution of the parameters in 
many {\it Fermi}--LAT blazars (as listed e.g. in Ghisellini et al. 2010b),
we conclude that the changes required to explain the observed 
(factor $\sim$1.5) variability are very small.

\section {Discussion and Conclusions}

The simultaneous observations of {\it NuSTAR} and {\it Swift}/XRT
revealed that the X-ray spectrum from $\sim$0.3 to 60 keV of both
sources are well described by a broken power-law model.
Both indices are very hard, with no requirement of 
absorption in excess of the Galactic one.
This broken power-law behavior is well reproduced by the model as due to the 
combination of two effects.
First, the electron energy distribution, below the cooling energy $\gamma_{\rm c}$,
retains the slope of the injected electrons (i.e. $\gamma^{-s_1}$), which is very hard
(see Table \ref{para}).
Second, what we observe as low-energy X-ray emission at the frequency $\nu_x$ 
is seen, in the comoving frame, as ultraviolet emission (i.e. $\nu_x^\prime=\nu_x(1+z)/\delta$), 
produced by the scattering between low-energy electrons and 
seed photons with frequencies smaller than $\nu_x^\prime$.
In both sources, the main contributions to the seed photons come
from the IR torus, the BLR, and the disk.
All these components have a corresponding radiation energy density 
peaking at some frequency, which is seen between the optical and the UV
in the comoving frame of the jet.
Inverse Compton scattered photons at, for example, $\nu^\prime \sim 10^{15}$ Hz  
can be produced using seed photons of frequencies smaller than $10^{15}$ Hz,
which do not correspond to the full energy density of the seed photons.
Scattered photons of higher energies, instead, can use the entire 
seed photon distribution.
In other words, there is a paucity of seed photons at low energies,
making the Inverse Compton spectrum harder below a few keV.
What is observed in the two blazars studied here can then be regarded as a proof
of the ``External Compton" process, where the major contribution to the
seed photons comes from radiation produced externally to the jet.
Furthermore, the spectrum between 0.3 and 60 keV is smooth, indicating the
absence of other emitting components.

Both blazars show variability, but not extreme variability.
This agrees with a relatively large source size, and a correspondingly
relatively large distance from the black hole.
The observed light crossing time predicted by our model is 
$t^{\rm obs}_{\rm var}\equiv (R/c) (1+z)/\delta$, corresponding to
$\sim$13 days for S5~0836+710 and to $\sim$10 days for PKS~2149--306.
This agrees with no variability observed within a single epoch observation,
and is consistent with the emission site being between the BLR 
and the torus ($R_{\rm BLR}<R_{\rm diss}<R_{\rm torus}$).
We note that, contrary to what we have observed, significant variability on shorter time-scales has been reported
for S5 0836+710 by Akyuz et al. (2013). In the framework of our modeling this is explained by assuming that the component
dominating the emission at a given time is not always located in the same place along the jet.
Sometimes it can be at a position closer to the black hole, and be more compact, giving rise to variability
on  shorter time-scales and to a slightly different SED.

We have shown that the observed moderate variability can be produced
by a rather small change in the injected power, break energy of the 
electron distribution and in the location of the emitting region.
The largest change is needed for PKS 2149--306, requiring a factor of two
change in the injected power.

The {\it NuSTAR} data at high energies, coupled with the $\gamma$-ray
data, allow us to determine where the high-energy hump of the SED peaks.
This is at $\sim$1 MeV (observed, thus $\sim$3 MeV, rest frame) for both sources. 
As already mentioned, this helps to determine the location of the emitting 
region, making us prefer an emitting region located between the BLR and the torus. 
We stress that this may not none thiecessarily always be the case 
since the emitting site can change, and sometimes it can be within the BLR.
For a given electron distribution, the high-frequency peak 
would shift to higher values, in this case.
The data of S5~0836+710 have been analysed also by Paliya (2015), who
fit the data using the same one single zone leptonic emission model.  
We find consistent results, but our black hole mass is slightly larger,
and our emitting region is just outside the BLR, while in Paliya et al. 
(2015) is inside. 
We find that while the parameters of Paliya et al. (2015) indeed
satisfactorily fit the overall SED and the average {\it NuSTAR}
spectrum, our fit accounts for the two states of the source in a better 
way, and in particular the two sets of {\it Swift}/XRT+{\it NuSTAR} data.

Both blazars have a black hole with mass exceeding $10^9 M_\odot$
and accretion disks emitting at about one-third of the Eddington rate.
The power that the jet expends for producing the radiation we see ($P_{\rm r}$ 
in Table \ref{powers})
is of the same order as $L_{\rm d}$ (note that the listed values of $P_{\rm r}$
refer to one jet only, and should be doubled).
$P_{\rm r}$ is a {\it lower limit} of the true jet power $P_{\rm jet}$. 
An estimate of $P_{\rm jet}$ can be derived assuming, following Nemmen et al. (2012),
that $P_{\rm jet}\sim 10 P_{\rm r}$.
Another estimate can be found assuming that there is one proton for each emitting
electron (this is the quantity $P_{\rm p}$ in Table \ref{powers}).
Either way, we are forced to conclude that the jet power exceeds the luminosity
of the accretion disk, thus suggesting that the jet is powered not only by accretion,
but also by the rotational energy of a spinning black hole,
as found by Ghisellini et al. (2014) for a large sample of blazars.

Finally, given the very hard spectrum detected by {\it NuSTAR}, and their high-energy peak
at $\sim$1--10 MeV, one can wonder if powerful blazars can significantly contribute to 
the X-ray background above its $\sim$30 keV peak.
While a complete study of this issue is still missing, there are preliminary results
by Draper \& Ballantyne (2009), Giommi (2011) and by Comastri \& Gilli (2011), suggesting that blazars
can contribute at the 10\% level at $\sim$100 keV, before becoming the dominant contributors
to the $\gamma$-ray background (Abdo et al., 2010; Ajello et al. 2015).
%
%
\vskip 0.5 cm
\section*{Acknowledgements}
We acknowledge financial support from the ASI-INAF grant I/037/12/0.
This work was supported under NASA Contract No. NNG08FD60C, and
made use of data from the {\it NuSTAR} mission, a project led by
the California Institute of Technology, managed by the Jet Propulsion
Laboratory, and funded by NASA. 
We thank the {\it NuSTAR} Operations, Software and Calibration teams for support 
with the execution and analysis of these observations.  
We also thank the {\it Swift} team for quickly approving and executing the requested
ToO observations.  
This research has made use of the {\it NuSTAR} Data Analysis Software
(NuSTARDAS) jointly developed by the ASI Science Data Center (ASDC,
Italy) and the California Institute of Technology (Caltech, USA).
The \textit{Fermi} LAT Collaboration acknowledges generous ongoing support
from a number of agencies and institutes that have supported both the
development and the operation of the LAT as well as scientific data analysis.
These include the National Aeronautics and Space Administration and the
Department of Energy in the United States, the Commissariat \`a l'Energie Atomique
and the Centre National de la Recherche Scientifique / Institut National de Physique
Nucl\'eaire et de Physique des Particules in France, the Agenzia Spaziale Italiana
and the Istituto Nazionale di Fisica Nucleare in Italy, the Ministry of Education,
Culture, Sports, Science and Technology (MEXT), High Energy Accelerator Research
Organization (KEK) and Japan Aerospace Exploration Agency (JAXA) in Japan, and
the K.~A.~Wallenberg Foundation, the Swedish Research Council and the Swedish
National Space Board in Sweden. Additional support for science analysis during the 
operations phase is gratefully acknowledged from the Istituto Nazionale di Astrofisica 
in Italy and the Centre National d'\'Etudes Spatiales in France.
Part of this work is based on archival data, software or on-line services 
provided by the ASI Data Center (ASDC).


\label{lastpage}
\end{document}